# Adsorption of fragrance capsules onto cellulose nano- and micro-cellulose fibers in presence of guar biopolymers

Evdokia K. Oikonomou and Jean-François Berret*

*Université de Paris, CNRS, Matière et systèmes complexes, 75013 Paris, France*

**Abstract** Fabric softeners are formulated to enhance textile softness and impart a pleasant scent. One of the most efficient technologies for controlled fragrance delivery onto fabrics involves encapsulating scent molecules in polymer capsules. Here, we investigate the adsorption of anionic fragrance capsules on cotton fabrics with the goal of reducing the reliance on palm-oil-derived surfactants. First, we employ 200 nm-cellulose nanocrystals (CNC) as a reliable model for cotton fibers. CNC enables us to explore interactions among various softener components, including surfactants, guar biopolymers, and fragrances, using physical chemistry techniques applied to bulk dispersions. The primary objective is to elucidate the role of surfactant vesicles, the primary ingredient in textile conditioners, in the association between fragrance capsules and cotton. Secondly, we examine the influence of biopolymers present in a newly developed, environmentally friendly softener on this association. Our findings demonstrate that anionic fragrance capsules are deposited onto cotton microfibers in the presence of either cationic surfactants or guar biopolymers, driven by electrostatic interactions. Scanning electron microscopy confirms capsule adsorption on textile fibers when these cationic ingredients are present. Understanding the interaction mechanisms between fragrance capsules and cotton fabrics, as well as the roles played by other softener components, can facilitate the design of more efficient and sustainable formulations.

***Keywords**: Fabric Softeners, Fragrance Capsules, Nanocellulose, Vesicles, Cotton*

*Corresponding author: Jean-François Berret
E-mail address: jean-francois.berret@u-paris.fr

## 1 - Introduction

Fabric softeners, also known as fabrics conditioners are a category of household products that have attracted consumer interest for their ability to make textiles softer, smoother and more pleasant smelling. [1-3]. Fabric conditioners are mainly composed of cationic double-tailed surfactants, with surfactant concentration around 10 wt. % for concentrated formulations. These surfactants are assembled in nano-and micro-sized vesicles in deionized water [4-8], whereas the vesicles are essential for the formulation colloidal stability and also for driving additives such as oil fragrances on the textile fibers [9-12]. Although conditioners have been used for decades, only recent studies have addressed the deposition and softening mechanisms. Pertaining to the interaction with fabrics, it is now recognized that the surfactant vesicles spontaneously adsorbed to fibers in the wet state [6,7,13,14]. Upon slow evaporation of water and drying, the vesicles break up to form a surfactant supported bilayer that coats the textile fibers [8]. The adsorbed surfactants are responsible for the softness effect via friction reduction associated with the sliding of the bilayers during fabric mechanical movement. There are however few physicochemical studies to date in the field of fragrance for softeners [15], and the knowledge about microscopic mechanisms is essentially empirical.

Aroma and fragrances are introduced in softener formulations both as essential oil (*i.e.* low molecular weight hydrophobic molecules) and in capsule form. Until recently, only fragrance oil was used. However, since oils do not adsorb directly to fabrics, a large proportion is lost as they are washed away after the softener addition cycle. A solution to this issue came from the encapsulation technology [16], permitting a long-lasting perfume activity due to a controlled fragrance release. Recently, a major effort has been devoted to the development of such micron-sized capsules [11,17-29]. Synthetic or natural polymers, inorganic materials and polymer-inorganic composites have been identified as chemical compounds for capsule scaffolds [3,18]. In the previous list, polymer-based nano- or micro-particles are among the most used due to their sizes, excellent mechanical properties and easy processing [9]. These capsules are designed to prolong fragrance perception by releasing the scent molecules when their envelope breaks by friction with the fabric [25,30]. This extends the fragrance activity on textiles for weeks, and in some cases the capsules are able to resist several wash cycles.

The most common polymer capsules used for fragrance encapsulation to date are melamine-formaldehyde resins due to their mechanical properties, low cost and simplicity of synthesis [19,20,23,25,29]. However, melamine-based capsules lack of biodegradability and they release formaldehyde. Alternatives such as formaldehyde-free melamine resins [23] or polysulfones, polyurea [22], natural polymers [21,26,27,31] and graft copolymers have been implemented [10]. Despite the progress already made in encapsulation field, the search for new coating materials and controlled fragrance release is still ongoing to meet the needs for prolonged aroma activity and limited capsules desorption from the textiles [32]. Additionally, reducing the laundry washing frequency will have a significant impact on reducing pollution from fabric microfibers [33]. In this regard, it should be noted that encapsulation itself is considered as a milestone towards sustainability and economical profit as it reduced the release of organic compounds in wastewater and the frequency of use of washing machine due to long-lasting perfume efficiency.

To accelerate the design of more eco-friendly formulations, it is crucial to comprehend the capsule adsorption and desorption mechanisms on and from the fabrics in relation to the physical-chemical properties of the whole formulation. Scanning electron and fluorescent microscopy have been used for instance to examine encapsulated fragrance adsorption on cotton fibers [17]. Other reports point out that the aroma molecules adsorption on cotton fibers is enhanced by cationic surfactants due to their solubilization within micelles or vesicles, and the electrostatic interaction of the cationic surfactants with the anionic cotton fibers [34,35]. However, to date, detailed studies on the deposition mechanism in correlation with the other ingredients of the softener are still lacking.

In the present work we report on a method to evaluate the deposition of fragrance capsules on cotton by using 200 nm long cellulose nanocrystals (CNC) as a model system. CNCs are rod-like nanoparticles derived from cotton or wood cellulose, well dispersed in aqueous media thanks due to their negative electrostatic charge and colloidal sizes [36,37]. CNCs thus allow us to mimic the interaction of capsules with cellulose in experiments using bulk dispersions. Softener formulations are indeed complex as they contain multiple chemical compounds, as well as additives (e.g. pigments, antimicrobial agents) in lesser proportions. In this context, it is important to unravel the role of each constituent on the deposition and softening mechanism and to this end experiments on nanocellulose enable us to assess the cross-effects and potential synergies between the different formulation components. For instance, we have previously applied this method to estimate the softener surfactant vesicles and polymers on cotton fibers [13,14,38,39]. Our goal here is to provide a straightforward protocol to quickly and simply identify the strength of interaction between the fragrance capsules and CNCs, and determine the role of the other softener ingredients in this interaction. Here, we use negatively charged melamine-based capsules [24] and we monitor their interaction on cellulose alone, in presence of a benchmark softener containing surfactant vesicles [1,2]. Furthermore, we harness a newly developed conditioner with a lower environmental impact due to the reduced esterquat surfactant content. This formulation designed by Solvay® comprises minor quantities of cationic and hydroxypropyl guar biopolymers (C-Guar and HP-Guar) which compensate the surfactant reduction on performance. Here we show that the improved fragrance delivery performance of such formulations [40,41] is due to the increased capsule deposition onto cotton in a mechanism that is driven by both the esterquat surfactant and cationic guar. We show that the anionic fragrance capsules by themselves do not interact with same charged cellulose but the cationic vesicles or the guar drive them on the cellulose fibers. The proposed method, combined with scanning electron microscopy, can be applied with sensory testing technologies to accelerate the design and production of more efficient and less environmentally toxic encapsulated fragrances in household products.

# 2 – Materials and Methods

## 2.1 - Materials

The quaternary surfactant, TEQ (abbreviated for ethanaminium 2-hydroxyN,N-bis(2-hydroxyethyl)-N-methyl-esters) with C16-18 aliphatic chains, was supplied by Solvay®. The chemical structure of TEQ is depicted in **Fig. 1a**. TEQ exhibits a gel-to-fluid transition at $T_M$ = 60°C due to molecular long-range order within the membrane [8,14]. **Fig. 1a** illustrates the presence of methyl sulfate anions as counterions linked to the quaternized amines in TEQ. Dynamic light scattering (DLS) studies revealed the presence of particles in the size range of 0.1 to 3 μm. A bimodal size distribution is observed for a TEQ dispersion at $c_{TEQ}$ = 0.1 wt. %, with peaks at approximately 140 nm and 800 nm, as shown in **Fig. 1a**. These particles have been previously identified as surfactant vesicles [14]. Cryogenic electron microscopy (Cryo-TEM) analysis further confirmed the presence of both unilamellar and multivesicular vesicles, indicated by red and blue arrows in **Fig. 1a**. Zetametry characterization indicated that TEQ vesicles carry a highly positive charge ($\zeta$ = +65 mV). **Table I** summarizes the various chemicals included in the topical softener formulation examined in this study, along with their respective abbreviations, concentrations, and molecular or colloidal states.

Cellulose nanocrystals used in this study were produced through a catalytic and selective oxidation process. Cotton linters supplied by Buckeye Cellulose Corporation® were hydrolyzed using the method described by Revol *et al.* [42]. This involved treating the cellulosic substrate with 65 wt./vol. % sulfuric acid at 63 °C for 30 minutes. The resulting suspensions underwent repeated centrifugations, dialysis against deionized water until neutrality, and were then sonicated for 4 minutes using a Branson B-12 sonifier equipped with a 3 mm microtip. Subsequently, the suspensions were filtered through 8 μm and then 1 μm cellulose nitrate membranes (Whatman). This process yielded a $c_{CNC}$ =2 wt. % aqueous stock suspension. CNC dispersions at 0.1 wt. % were characterized using cryo-TEM and DLS, with results presented in **Fig. 1b**. Cryo-TEM revealed well-dispersed particles in the form of laths with an average length of 180 ± 30 nm, width of 17 ± 4 nm, and thickness of 7 ± 2 nm. DLS confirmed their size, centered at 120 nm. Deionized water was obtained using a Millipore Milli-Q Water system.
The polysaccharide polymer, cationic guar (C-Guar), was synthesized by Solvay® through the modification of natural guar gum with cationic groups (**Fig. 1c**), resulting in a positively charged copolymer ($\zeta$ = +30 mV). The obtained C-Guar had a molecular weight of $0.5 \times 10^6$ g mol$^{-1}$ and a degree of substitution of approximately 0.1. DLS size distribution, shown in **Fig. 1c** for $c_{C-Guar}$ = 0.02 wt. % exhibited a single peak around 200 nm. This hydrodynamic diameter was higher than that estimated for polysaccharides with similar molecular weights [43], indicating potential chain association in water, forming hydrocolloid particles. C-Guar was selected for its known properties in providing smoothness and conditioning to hair [44,45]. Studies have demonstrated its strong adsorption on cotton [40,41], compensating for the TEQ reduction in the fabric softener formulation in terms of performance.

We employed melamine-formaldehyde capsules containing fragrance as model capsules [17,19,20,25], which were supplied by Solvay®. We will refer to these fragrance capsules as CAPS in the following sections. Aqueous CAPS dispersions at a concentration of 1 wt. % were subjected to characterization using phase-contrast optical microscopy, and the results are presented in **Fig. 1c** at two different magnifications. The capsules are indicated by arrows. The size distribution in the first image is illustrated in **Fig. 1c**, with the peak centered at 1 μm.

| Chemicals | Abbreviation | Concentration (wt. %) | Molecular or colloidal State |
|---|---|---|---|
| Esterquat surfactant | TEQ | 4.0 | Multivesicular vesicles |
| Cationic guar | C-Guar | 0.2 | Branched polymers |
| Hydroxypropyl guar | HP-Guar | 0.4 | Branched polymers |
| Essential oil | Essential oil | 1.0 | Hydrophobic molecules |
| Fragrance capsules | CAPS | 1.0 | Microspheres |
| Additives (pigment, antimicrobial agents, salt) | Additives | < 1.0 | Ions, molecules |

***Table I*** *: List of chemicals present in topical formulations of fabric softeners, together with abbreviation, concentration and a description of the molecular or colloidal state.*

## 2.2 - Sample preparation

Aqueous TEQ dispersions were prepared by melting the neat TEQ paste and adding it to deionized water at 60 °C. The pH was adjusted to 4.5. TEQ/C-Guar formulations containing $c_{TEQ}$ = 6 wt. % and $c_{C-Guar}$ = 0.4 wt. % were prepared by first dispersing the guar polymer in 60 °C deionized water, adjusting the pH to 4.5, and then adding the melted TEQ surfactant in the final step. For DLS, $\zeta$-potential, and microscopy studies, the dispersions were diluted by adding the appropriate volume of deionized water at a temperature of 25 °C. Aqueous suspensions of CAPS were prepared by dilution with the desired volume of deionized water. For DLS experiments, the CAPS suspensions were filtered using a paper filter. The concentration of the filtered dispersion was determined by evaporation and weighing, resulting in a concentration of $c_{CAPS}$ = 0.5 wt.%. CNC aqueous dispersions were obtained by diluting the $c_{CNC}$ = 2 wt. % stock dispersion [46] with deionized water and subsequently adjusting the pH to 4.5.

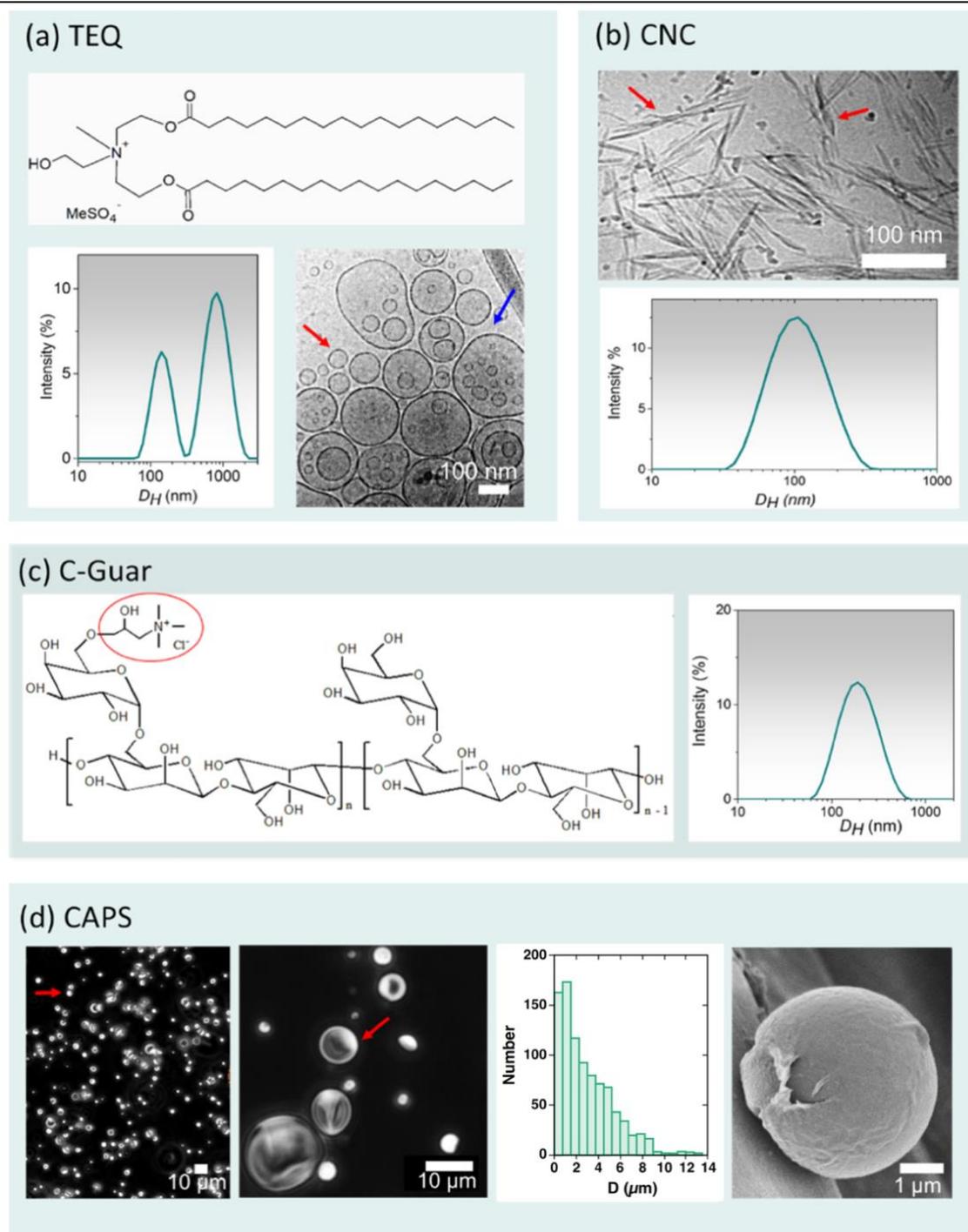

**Figure 1**: **a)** TEQ double-tailed surfactant chemical structure (upper panel), size distribution obtained by dynamic light scattering (lower-left panel) and cryo-transmission electron microscopy of a $c_{TEQ}$ = 0.1 wt. % TEQ dispersion (lower-right panel). **b)** Cellulose nanocrystal cryo-TEM image (upper panel) and size distribution obtained by DLS for $c_{CNC}$ = 0.1 wt. % aqueous dispersion (lower panel). **c)** Chemical structure (left panel) and size distribution obtained by dynamic light scattering for a $c_{C-Guar}$ = 0.2 wt. % dispersion (right panel). **d)** Representative images of phase contrast optical microscopy at magnifications ×20 and ×40 of a CAPS aqueous dispersion at $c_{CAPS}$ = 0.1 wt. % (left panels). The fragrance capsules are shown by the arrows. The right center panel displays the CAPS size distribution

obtained from treatment of the microscopy images ×20. A scanning electron microscopy image of a fragrance capsule is provided in the right panel.

| Compounds | $D_H$ (nm) | pdi | $\zeta$-potential (mV) |
|---|---|---|---|
| Surfactant (TEQ) | 220 | 0.45 | +47 |
| Cationic guar (C-Guar) | 250 | 0.80 | +30 |
| Fragrance capsules (CAPS) | 310 | 0.30 | -13 |
| Cellulose nanocrystal (CNC) | 120 | 0.21 | -38 |

*Table II: List of hydrodynamic size, dispersity index and $\zeta$-potential of the chemical compounds incorporated in the softener formulation (lines 1 to 3) as well as the same characteristics for cellulose nanocrystals (CNCs).*

## 2.3 - Mixing protocol

We examined the interactions among CNC, CAPS, C-Guar, TEQ, or their mixtures using the direct mixing formulation pathway [14,47,48]. Aqueous dispersions of the above-mentioned materials were prepared at concentration $c$ = 0.005 or 0.01 wt. % and then mixed at different ratios, noted $X = c_A/c_B$ where $c_A$ and $c_B$ are the A and B weight concentrations in the mixed dispersions. The total concentration remained constant at $c = c_A + c_B$. In the following, A and B may be designated by one of the following abbreviations: TEQ, C-Guar, CAPS, CNC. The mixed dispersions were prepared in the range $X$ = 10⁻³ -10³ at room temperature ($T$ = 25 °C). By convention, we assign the neat TEQ, C-Guar, CAPS and CNC dispersions to the values of $X$ = 0.001 or $X$ = 1000 [8,47]. Following mixing, the dispersions were rapidly stirred, allowed to equilibrate for 5 minutes, and subsequently characterized using light scattering, $\zeta$-potential measurements, and optical microscopy.

## 2.4 - Dynamic light scattering

The scattering intensity $I_S$ and hydrodynamic diameter $D_H$ were determined using the Zetasizer NanoZS spectrometer (Malvern Instruments, Worcestershire, UK). A 4 mW He−Ne laser beam ($\lambda$ = 633 nm) is used to illuminate the sample dispersion, and the scattered intensity is collected at a scattering angle of 173°. The second-order autocorrelation function $g^{(2)}(t)$ is analyzed using the cumulant and CONTIN algorithms to determine the average diffusion coefficient $D_C$ of the scatterers [28]. The hydrodynamic diameter is obtained from the Stokes-Einstein relation, $D_H = k_B T/3\pi\eta D_C$, where $k_B$ is the Boltzmann constant, $T$ the temperature and $\eta$ the solvent viscosity (here 0.89 mPa s). Measurements were performed in triplicate at 25 °C after an equilibration time of 120 s. It is worth noting that CNCs are colloids with differing translational diffusion constants parallel and perpendicular to their main axis. Following the convention established in the literature for CNCs [36], we refer to this parameter as the apparent hydrodynamic diameter, $D_H^{app}$.

## 2.5 - $\zeta$-potential

Laser Doppler velocimetry (Zetasizer, Malvern Instruments, Worcestershire, UK) using the phase analysis light scattering mode and a detection at an angle of 16 degrees was performed to determine the electrophoretic mobility and $\zeta$-potential of the different dispersions and suspensions studied. Measurements were performed in triplicate at 25 °C, after 120 s of thermal equilibration.

### 2.6 - Phase-contrast optical microscopy

Images were captured using an IX73 inverted microscope from Olympus, equipped with a 60× objective lens. The acquisition system included an Exi Blue camera from QImaging and Metaview software from Universal Imaging Inc. For sample preparation, 8 microliters of the CAPS dispersion were deposited on a glass plate and sealed within a Gene Frame dual adhesive system (Abgene/Advanced Biotech). Subsequently, the acquired images were digitized and processed using ImageJ software and its associated plugins ([http://rsbweb.nih.gov/ij/](http://rsbweb.nih.gov/ij/)).

### 2.7- Cryogenic transmission electron microscopy (Cryo-TEM)

A volume of seven microliters of the samples was dispensed onto lacey carbon-coated 200 mesh grids (Ted Pella). The excess liquid was gently removed by blotting the grid three times using filter paper using a VitrobotTM (FEI). Subsequently, the grid was rapidly plunged into liquid ethane and cooled using liquid nitrogen to prevent the crystallization of the aqueous phase. The membrane-coated grid was then transferred into the vacuum column of a TEM microscope (JEOL 1400 operating at 120 kV), where it was maintained at a temperature of liquid nitrogen using a cryo-holder (Gatan). Images were recorded at magnifications ranging from 3000× to 40000×, using a 2k-2k Ultrascan camera (Gatan).

### 2.8 - Scanning electron microscopy (SEM)

Cotton fabrics measuring 5 x 5 cm were used as a substrate to examine the application of the softener to the fibers. These fabrics were made from cotton yarns with a diameter 300 μm, themselves made from fibers with diameters between 10 and 20 nm. Before treatment with the softener formulation, the fabrics were immersed in Milli-Q water for 10 minutes, followed by drying at 35°C. These fabrics were then immersed in a 100 ml fabric softener solution for 10 minutes with continuous agitation. After this treatment, the fabrics were placed in a horizontal, non-contact position and allowed to air-dry at 35°C for one hour. The cotton fabrics studied by scanning electron microscopy (SEM) were spread on a silicon chip cleaned using a Gatan plasma cleaner to make them hydrophilic and covered by a silicon nitride amorphous film [8]. Experiments were realized on a ZEISS Gemini SEM 360 equipped with an Oxford Instruments Ultim Max 170 mm² detector (ITODYS Laboratoire, Paris). Samples were stuck in the sample holder using conductive double-sided adhesive tapes. SEM images were obtained by Inlens SE detector (In Column) at 5 kV accelerating voltage. The magnification achieved under these conditions was between X100 and X40,000.

# 3 - Results and Discussion

## 3.1 - Interaction of fragrance capsules with cationic surfactants and C-Guar polymers

In this section we investigate the interaction of the fragrance capsules with the main softener components, the TEQ esterquat surfactant and the C-guar cationic polysaccharides. For this, we examine the phase behavior of TEQ/CAPS/water ternary dispersions by applying the continuous variation method originally proposed by Paul Job in 1928 [49,50] and later adapted by us for various techniques, including for light, small-angle neutron and X-ray scattering [48,51]. The Job scattering plot in **Fig. 2a** displays the hydrodynamic diameter $D_H$ obtained by DLS for TEQ/CAPS mixed dispersions at the concentration $c = 0.01$ wt. % as a function of the ratio $X = c_{TEQ}/c_{CAPS}$. Here $c_{TEQ}(X)$ and $c_{CAPS}(X)$ denote the TEQ and CAPS weight concentrations at a given $X$-value, respectively. In **Fig. 2a** the total concentration $c = c_{TEQ} + c_{CAPS}$ is constant for all dispersions and remains in the dilute regime, a prerequisite for light scattering [28,52]. It is found that the $D_H(X)$ exhibits a broad peak at the critical mixing ratio $X_C = 0.2$ (arrow). The $D_H(X)$ for the pure dispersions at $X = 10^{-3}$ (CAPS) and $X = 10^3$ (TEQ) are found to be 310 and 220 nm, respectively (**Table II**). The increased $D_H(X)$ is attributed to the formation of mixed TEQ/CAPS aggregates resulting from electrostatic driven interaction. The formation of aggregates was further confirmed by optical microscopy (data not shown). The $\zeta$-potential was also measured for mixed TEQ/CAPS dispersions. In **Fig. 2b**, it can be seen that the $\zeta$-potential gradually changes from negative values ($\zeta_{CAPS}$ = - 13 mV) to positive values ($\zeta_{TEQ}$ = + 47 mV) with increasing $X$. The point of zero charge is found around $X$ = 0.5, in good agreement with the maximum observed in the $D_H(X)$ behavior. Such a pattern of mixing of oppositely charged species has been found in a variety of colloidal systems, including synthetic and biological polymers, phospholipid vesicles and surfactants [53-58]. These outcomes support the assumption that the two components interact *via* electrostatic interaction. For ease of comparison among different results, this work systematically presents Job scattering plots in which negatively charged colloidal dispersions are depicted on the left-hand side, against a light grey background, while positively charged dispersions are displayed on the right-hand side, against a dark gray background. As previously mentioned, the critical value $X_C$ here corresponds to the charge stoichiometry that delineates the boundary between the negative and positive $\zeta$-potential domains [57].

Strong interaction and aggregate formation were also monitored for the C-Guar/CAPS mixed dispersions. A broad maximum in the Job scattering plot is observed in **Fig. 2c** at $X_C$ = 1 (arrow). The $\zeta$-potential experiments again show a progressive shift, again from negative values ($\zeta_{CAPS}$ = - 13 mV) to positive values ($\zeta_{C-Guar}$ = +30 mV) with increasing mixing ratio. The point of zero charge is found at $X$ ~ 0.5 in **Fig. 2d**, in agreement with the position of the maximum in $D_H(X)$. In both TEQ/CAPS and C-Guar/CAPS ternary dispersions, we observe an electrostatic co-assembly involving a process of titration and charge neutralization [59]. By knowing the electrostatic charge densities of the various components in the topical formulation (**Table I**), the continuous variation method is amenable for determining the charge stoichiometry. An important outcome here is that the fragrance capsules interact with both cationic components present in the softener formulation, and it is anticipated that these multiple interaction plays a role for their deposition on the cotton fibers. Finally, it is noted that in the Solvay® softener formulations, the TEQ/CAPS and C-Guar/CAPS mixing ratios are equal to $c_{TEQ}/c_{CAPS}$ = 4 and $c_{C-Guar}/c_{CAPS}$ = 0.2 (**Table I**), *i.e.* within the $D_H(X)$ interaction peaks seen in the Job scattering plots. This later result indicates that in actual formulations, the fragrance capsules are co-assembled with the cationic species TEQ and C-Guar, which prevent their sedimentation or creaming and increases the formulation stability.

### 3.2 - Impact of esterquat surfactant on fragrance capsule/nanocellulose interaction

Here, we search for synergistic effects between the components of the Solvay® softener formulation and encapsulated fragrances on their deposition on cotton fibers. Previous report by He *et al.* [11] has shown that melamine-based capsules interact with cellulose in the presence of salt, suggesting a deposition driven by electrostatic screening effect. For out part, we have developed a strategy that utilizes cellulose nanocrystals to quantify the interactions between softener components and cellulose materials. This approach has demonstrated a strong correlation between results obtained in bulk dispersions with CNCs and those on cellulose using cotton fibers [8,14]. Here, we use 200 nm long cellulose nanocrystals (**Fig. 1b**) as a cotton model to address the interaction of encapsulated fragrances on cellulosic substrates.

**Fig. 3a** displays the result of the continuous variation method applied to the CAPS/CNC/water ternary system. Here, the static light scattering intensity is displayed as a function of the mixing ratio $X = c_{CAPS}/c_{CNC}$ at a total concentration $c_{CAPS} + c_{CNC}$ = 0.01 wt. %. Of note, the samples were prepared using the same protocol as in **Figs. 2**. In **Fig. 3a** we observe that the measured intensity $I_S(X)$ increases progressively with $X$, and remains identical to the one calculated assuming that both CAPS and CNCs are noninteracting species (continuous line in green). This reference was obtained assuming that $I_S(X)$ is the sum of the scattered intensities of neat CAPS and CNC dispersions weighted by their respective concentrations. The result in **Fig. 3a** is direct evidence that both negatively charged CAPS and CNCs do not interact, but remain disperse in solution.

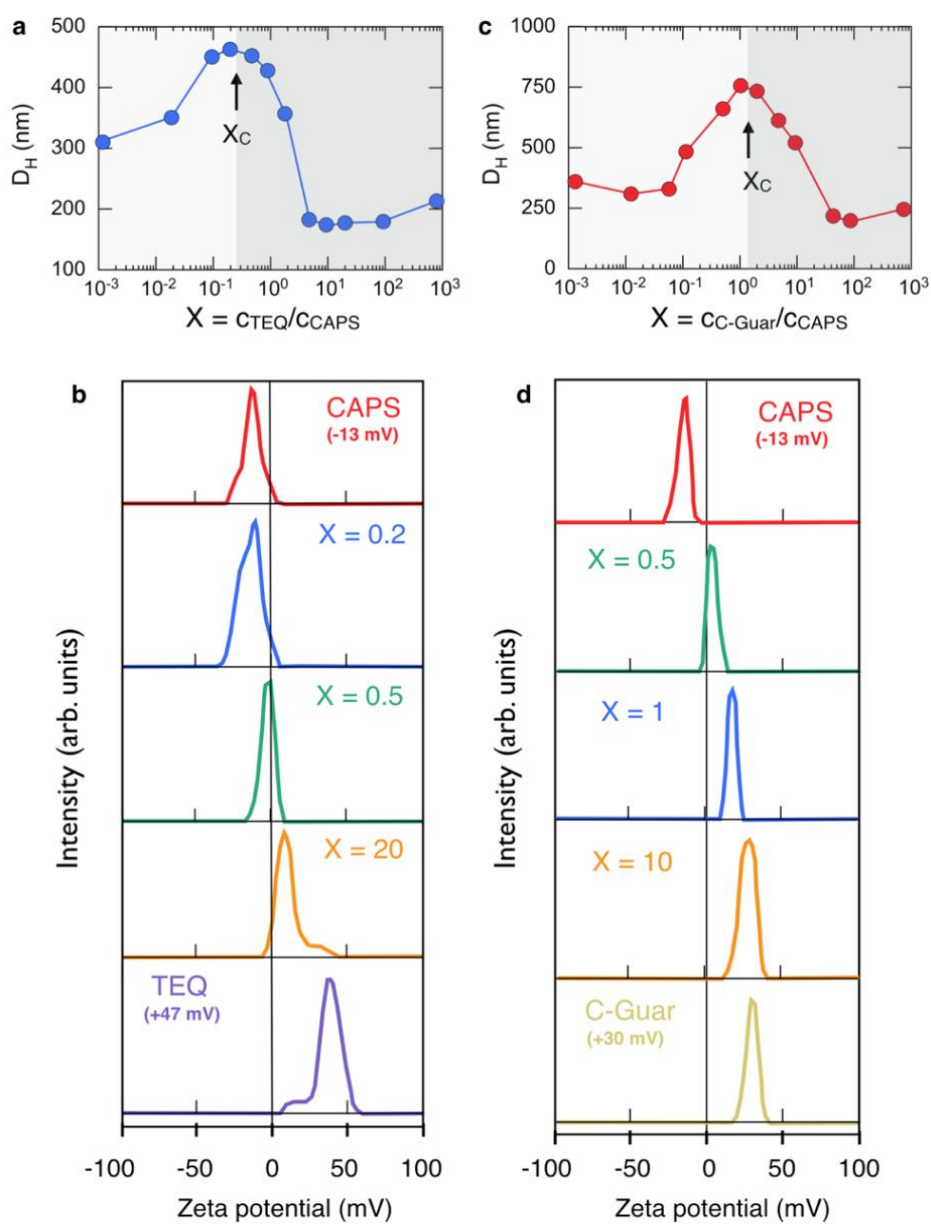

**Figure 2:** **a)** *Hydrodynamic diameters $D_H(X)$ and **b)** $\zeta$-potentials obtained for TEQ/CAPS mixed dispersions as a function of the mixing ratio $X$.* **c)** *and **d)** Same as in a) and b) for C-Guar/CAPS mixed dispersions. Abbreviations TEQ, CAPS and C-Guar denote the esterquat surfactant, the fragrance capsules and the cationic polysaccharide, respectively. The total concentration of the dispersions shown here is c = 0.01 wt. %. In **a)** and **c)**, the shaded areas apart from $X = X_C$ delineate the domains of negatively (left-hand side) and positively (right-hand side) charged colloids, respectively.*

**Fig. 3b** exhibits the scattered intensity $I_S(X)$ obtained for TEQ/CAPS mixed dispersions, the surfactants and the capsules being of opposite charge (**Table II**). The values presented here were recorded on the same samples than the hydrodynamic diameter in **Fig. 2a**. As for $D_H(X)$, the scattered intensity is found to pass through a maximum around $X_C$ = 0.5. It is in addition larger than the prediction for a non-interacting system (continuous line), confirming that TEQ and CAPS co-assemble upon mixing and form large aggregates. Similar results were obtained with TEQ and CNCs (**Fig. 3c**), although the effect here is weaker than in the previous case. It should be noted that for **Figs. 3a, 3b and 3c**, the experimental conditions were the same in terms of concentration, pH and temperature, and from their comparison it can be concluded that TEQ and CAPS exhibit the strongest interaction strength [14,48,60].

To verify the synergistic effect of TEQs and CAPS on the deposition of fragrances on cellulose, we then study TEQ/CAPS mixed dispersion with cellulose nanocrystals. The TEQ/CAPS concentration ratio is set to $X = c_{TEQ}/c_{CAPS}$ = 4, which corresponds to the actual ratio of the Solvay® formulation. To perform light scattering experiments and minimize sample absorbance however, the total TEQ and CAPS concentration is reduced with respect to that of the Solvay® formulation, and is set here at 0.0.1% by weight. In **Fig. 3d**, the scattered intensity $I_S(X)$ of the mixed dispersions deviates from the expected values for a non-interacting system (continuous line in green), indicating electrostatic driven co-assembly. This data suggests that the fragrance capsules also assemble with the cellulose nanocrystals through their prior association with the surfactants. Indeed, with a mixing ratio $X = 4$ between TEQ and CAPS, the structures are positively charged (**Fig. 2b**), and can interact with the negative CNCs. To our knowledge, such a result has not been reported yet in the context of fabric conditioners. This confirms that CNC can be used as a cellulose model to assess the interaction of different softener components, and to comprehend the role of these components on the conditioner deposition.

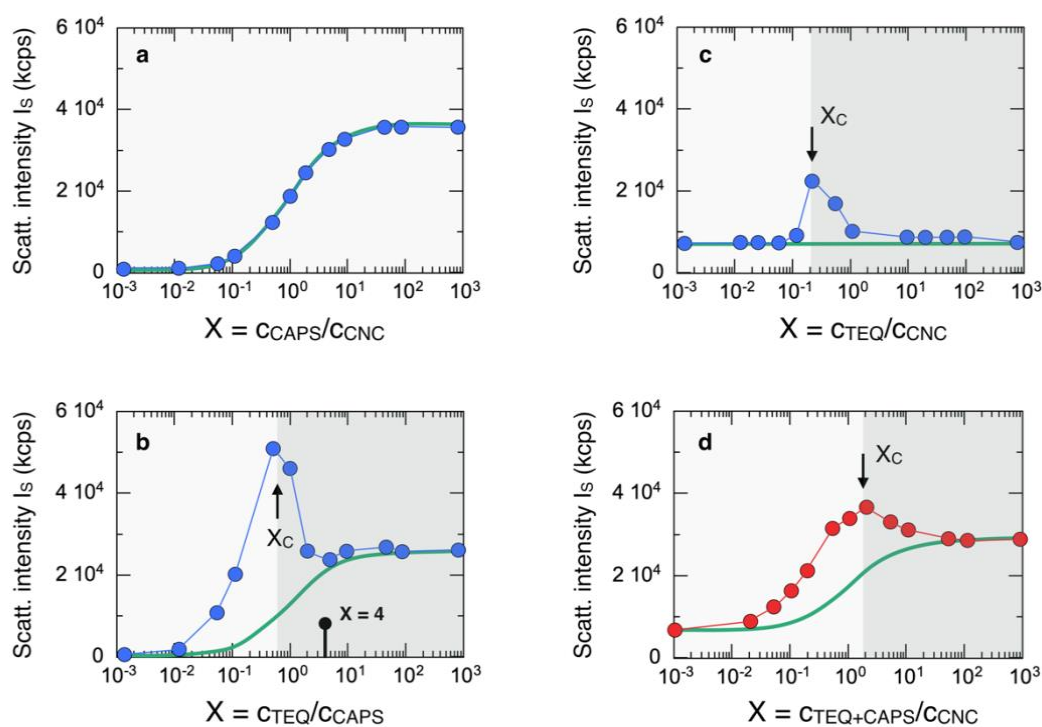

***Figure 3***: *Job scattering plots showing the static light scattering intensity of mixed CAPS/CNC **(a)**, TEQ/CAPS **(b)**, TEQ/CNC **(c)** and (TEQ+CAPS X=4)/CNC **(d)** dispersions as a function of the mixing ratio*

$X$. The continuous lines in green are calculated from the non-interacting model. The total concentration is constant and fixed at $c = 0.01$ wt. % for all samples. For pairs of opposite charge colloids (b-d), a co-assembly process is observed, with the formation of micron-sized structures and enhanced scattering. In b-d), the shaded areas apart from $X = X_C$ delineate the domains of negatively (left-hand side) and positively (right-hand side) charged colloids, respectively. The light gray back color in (a) indicates that the colloids remain negatively charged, whatever the whatever the $X$-values.

## 3.3 - Impact of cationic guar biopolymers on fragrance capsule/nanocellulose interaction

In this section, we replicate the continuous variation method experiments, substituting TEQ esterquat with G-Guar biopolymers. Our objective is to evaluate the synergistic effects of guar on the interaction between fragrance capsules and cellulose nanocrystals. We first evaluate separately the interaction between the cationic guars with CNCs (**Fig. 4a**) and with CAPS (**Fig. 4b**). In both cases we find characteristic patterns already found in **Figs. 2 and 3**, *i.e.* a strong increase of the scattered intensity with increasing mixing ratio and a sharp decrease beyond $X_C$. As the C-guars possess opposite charges compared to CNCs and CAPS (**Table II**), their mixture results in a pronounced electrostatic interaction among these distinct colloids, consequently yielding the observed scattering peak [38].

**Fig. 4c** shows the data obtained for the C-Guar/CAPS mixed dispersions at $X = 5$ with the cellulose nanocrystals. At this mixing ratio, the C-Guar/CAPS dispersion lies within the hydrodynamic diameter and scattering intensity peaks (**Fig. 2b** and **Fig. 4b**, respectively), and contains mixed aggregates. Upon CNC addition, micron-sized aggregates form at intermediate mixing ratios and precipitate in the range $X = 0.2 - 1.5$. The aggregates are significantly larger than those formed in the CAPS/CNC or C-Guar/CNC mixed dispersions, reaching sizes of several microns, despite having the same total active concentration. Here we see the same effect as with TEQ surfactants in **Fig. 3**, but the effect is enhanced with guar biopolymers, as indicated by the wide precipitation zone observed.

The results of **Figs. 3 and 4** suggest that although the fragrance capsules do not interact with the cellulose nanocrystals, the presence of cationic species such as TEQ surfactant or guar biopolymers favor this interaction. For this, however, it is necessary to formulate the set of three compounds in an appropriate way, so that the co-assembled structures are always positive, or neutral as it is made in the Solvay® topical formulation. In our previous studies [8,39], we have seen that guar biopolymers allow a reduction of esterquat by half (from 10 to 5 wt. %) without altering the softening properties. At the same time the yield stress property of the formulation is preserved, guaranteeing colloidal stability. We show here that cationic polymers more than compensate for the reduction in surfactant in terms of fragrance deposition, a quite remarkable result in the production of household products with reduced environmental impact.

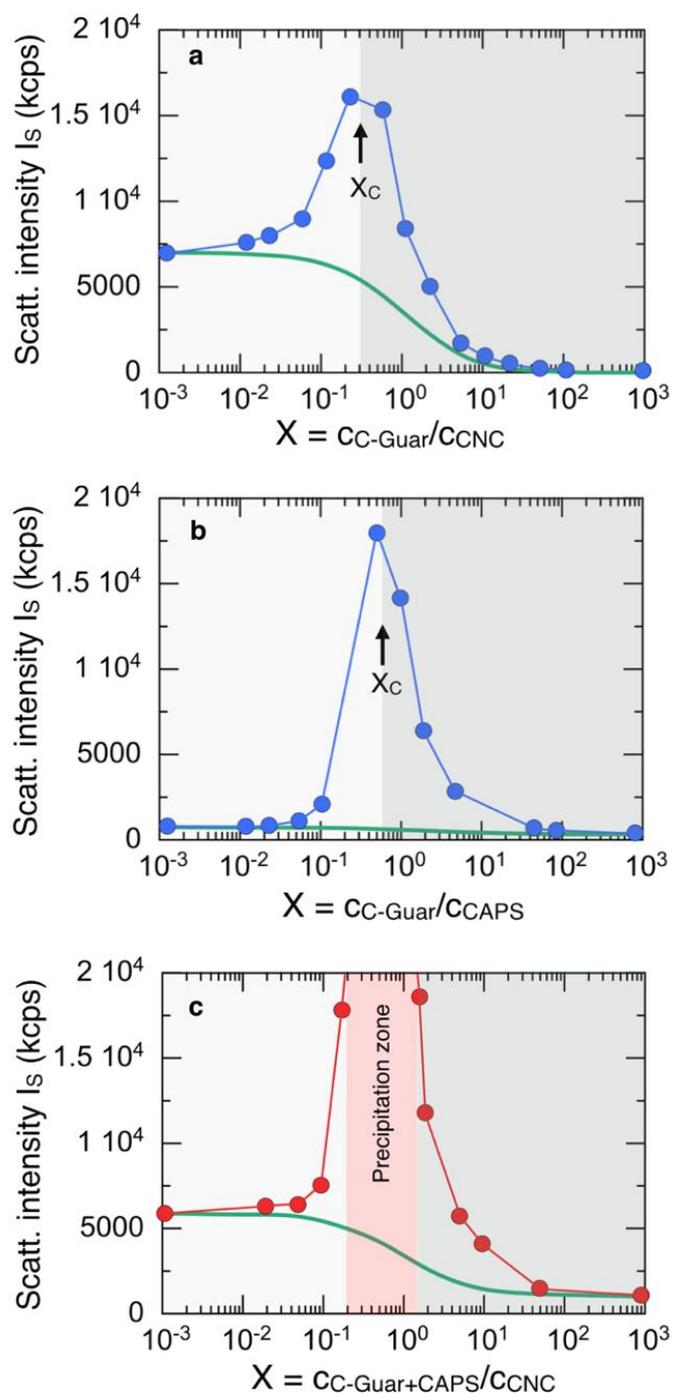

**Figure 4**: *Scattered intensity of mixed (a) C-Guar/CNC, (b) C-Guar/CAPS, (c) (CAPS+C-Guar X=5)/CNC as a function of the mixing ratio $X$. The continuous line in orange is calculated from the non-interacting model. The total concentration is constant and fixed at c = 0.01 wt. % for all mixtures. As in Figs. 2 and 3, the shaded areas apart from $X = X_C$ or apart from the precipitation zone delineate the domains of negatively (left-hand side) and positively (right-hand side) charged colloids, respectively.*

## 3.4 - Adsorption of fragrance capsules on cotton microfibers

In this section, we study the deposition of fragrance capsules on cotton fabric patches in the presence of TEQ, C-Guar or both. Centimeter large patches of fabric were immersed in the TEQ/CAPS or TEQ/C-Guar/CAPS dispersions at concentrations $c$ = 0.1 and 1 wt. % for 10 min under mild stirring and subsequently dried at 35°C for an hour. These concentrations were chosen based on our previous fluorescence and SEM studies for the optimal imaging of surfactant vesicles and polymers [8]. **Fig. 5a** displays a representative SEM image of untreated woven cotton fabric, comprising two sets of 250 μm spun yarns interlaced at right angles [61]. These yarns consist of interlocked cotton fibers separated by large voids (indicated by arrows). The cotton fibers appear as elongated cylinders with diameters ranging from 10 to 20 μm and exhibit crenulations and convolutions on their surfaces, as depicted in **Fig.5b** [62]. These fibers primarily consist of numerous fine threads of cellulose called fibrils (**Fig.5c**), which spiral around the fiber axis at an angle of approximately 70 degrees. In **Figs. 5d-f** representative SEM images of cotton fabrics treated with TEQ/CAPS X = 5 are presented. There are observable differences in the appearance of the cotton fibers compared to **Figs. 5b** and **5c** [8]. The fibrils mentioned earlier are no longer visible, indicating the presence of a TEQ surfactant supported bilayers [8]. The adsorption of capsules on cotton fibers is also easily recognized, their morphology and size corresponding to the optical microscopy results obtained in bulk suspensions (Materials and Methods section). Most of deposited capsules appear aggregated in large clusters, likely due to the combined effect of electrostatic interaction with vesicles and the relatively high concentrations used in the experiment. It is known that for electrostatic interactions, the size and morphology of co-assembled aggregates depend on the initial mixing concentration [47,57,63]. Furthermore, TEQ supported bilayers appear to wrap these CAPS clusters, as shown in **Fig. 5d**. In **Fig. 5e**, a single fragrance capsule is adsorbed onto the cotton fiber.

Close inspection of this SEM image reveals that a film deposited on the fiber extends towards the capsule, appearing to form a physical bond with it. As there are no other components in the formulation tested, it is reasonable to conclude that this film of some ten nanometers in thickness comes from the TEQ surfactants organized in the form of supporting bilayers. These observations confirm our earlier hypothesis regarding the synergy between TEQ and CAPS on cotton adsorption: TEQ is adsorbed on the capsules in the formulation, effectively wrapping them with a cationic film in the dried state, facilitating their deposition on anionic cellulose fibers. When C-Guar is introduced into the formulation (**Fig. 5g-i**), the deposition of biopolymers around the capsules is characterized by visible polymer-like threads and films (arrows). Hence, it is presumed that due to the strong interaction between C-Guar and CAPS, most of the guar has already been adsorbed onto them, even before fabric immersion in the treatment formulation. These C-Guar/CAPS aggregates are subsequently deposited on the fabrics. These findings support our earlier views that C-Guar contributes to CAPS adsorption onto the fibers. From an application point of view, it is worth recalling that in Solvay® patents, fabrics treated with guar containing conditioners were found to exhibit an improve fragrance performance [40,41]. This improvement can now be explained, and attributed to the significant role of C-Guar in fragrance adsorption.

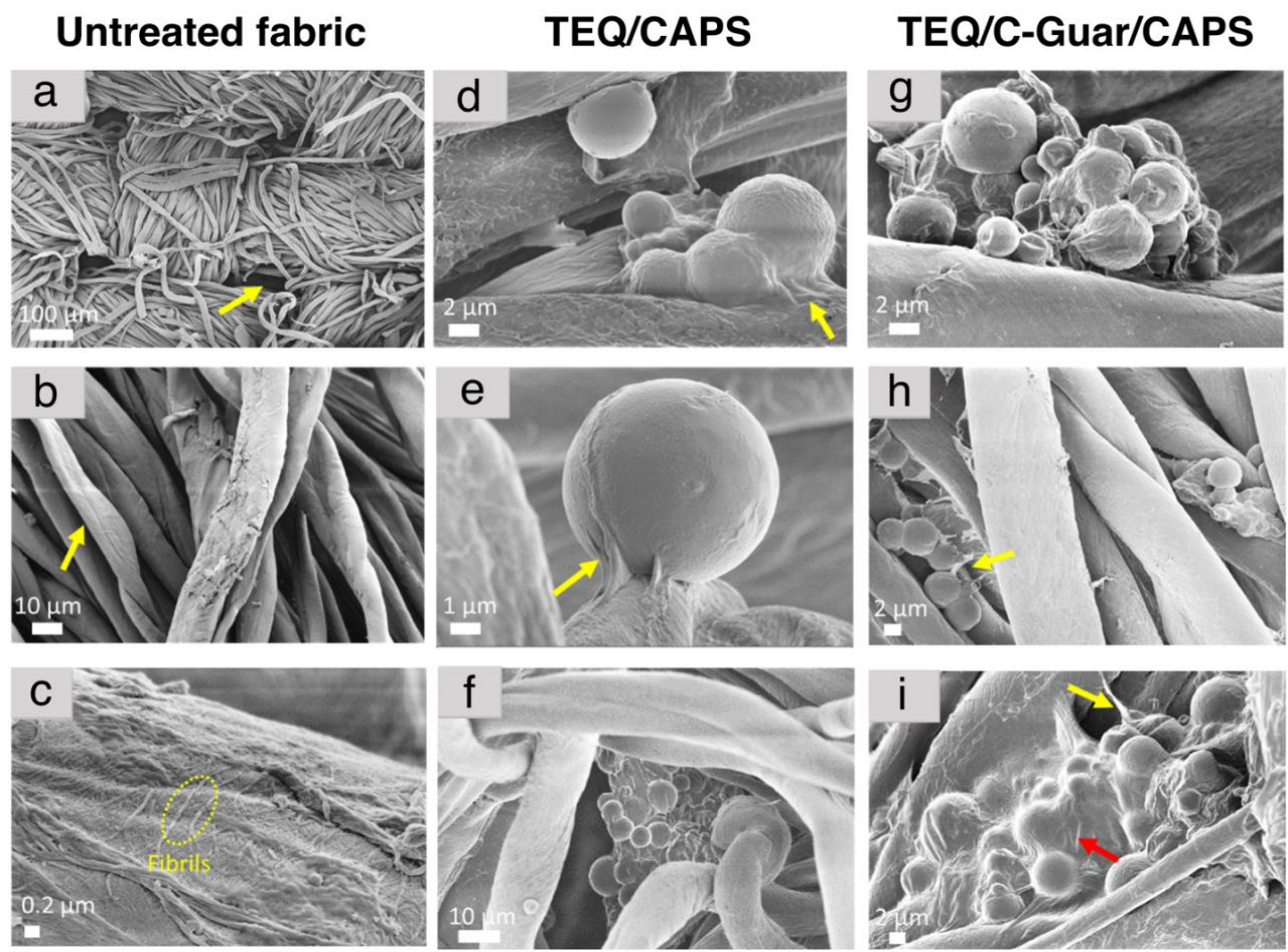

***Figure 5 :*** *Representative scanning electron microscopy images of cotton fabrics treated under the following conditions: **a) – c)** deionized water. **d) – e)** a mixed esterquat surfactant (TEQ) and fragrance capsule (CAPS) dispersion at concentrations $c_{TEQ}$ = 0.1 wt. % and $c_{CAPS}$ = 0.02 wt. % in the same ratio as in the Solvay® formulation (X = 5). **f)** a TEQ/CAPS dispersion at concentrations $c_{TEQ}$ = 1 wt. % and $c_{CAPS}$ = 0.2 wt. %. **g) – h)** a TEQ/C-Guar/CAPS dispersion at concentrations $c_{TEQ}$ = 0.1 wt. %, $c_{C-Guar}$ = 0.005 wt. % and $c_{CAPS}$ = 0.02 wt. %. **i)** a TEQ/C-Guar/CAPs dispersion at concentrations $c_{TEQ}$ = 1 wt. %, $c_{C-Guar}$ = 0.05 wt. % and $c_{CAPS}$ = 0.2 wt. %.*

## 4 - Conclusion

In this study, we investigate the deposition of the key ingredients present in topical fabric softeners onto cotton fibers. Although fabric softeners have been used for decades, it is only recently that efforts have focused on the mechanisms of deposition and softening, primarily to develop more environmentally-friendly formulations and reduce surfactant content. The formulations under investigation, developed in collaboration with Solvay® consist of cationic surfactants (TEQ), guar-based biopolymers (C-Guar), and micron-sized fragrance capsules. These capsules play a crucial role in modern fabric softeners to enhance textile softness and impart a pleasant scent. Fabric conditioners are by design complex dispersions where each ingredient has a specific function regarding the stability, delivery and softening mechanism. This study aims to unravel these mechanisms and explore potential cross-effects and synergies among the ingredients during deposition on cotton. First, we employ cellulose nanocrystals (CNCs) as model substrates, known for their similarities to cotton fiber surface physicochemistry, to assess interaction strengths through light scattering techniques, following a continuous variation method protocol [49,50]. Our key discovery is that deposition of surfactant and guar biopolymers occurs via electrostatic interactions between oppositely charged species. Positively charged TEQs and C-Guars exhibit strong interactions with negatively charged cellulose nano- and microfibers. The challenge arises when it comes to fragrance capsules, here made from melamine-formaldehyde resins, which are negatively charged and therefore not expected to adsorb on the cotton. However, our findings reveal that within the Solvay® formulation, fragrance capsules form clusters through interactions with both surfactant vesicles and guar biopolymers. Importantly, these clusters, positively charged, exhibit robust adsorption onto cotton fibers. This adsorption could be observed directly by scanning electron microscopy, which show supported surfactant bilayers wrapping the capsules, both being localized at the cotton fiber surface. In conclusion, this study provides crucial insights into the intricate interactions among surfactants, guar biopolymers, fragrance capsules and cellulosic surfaces. These insights lay the foundation for the development of environmentally-friendly formulations with enhanced fragrance delivery and improved softening properties in household products, contributing to sustainability and product efficiency.

## Acknowledgements

The authors thank Cristobal Galder and Nikolay Christov from Solvay® for fruitful discussions, support and advice. We would also like to thank Laurent Heux of the Centre de Recherches sur les Macromolécules Végétales (Grenoble, France) for helpful discussions and for providing us with the cellulose nanocrystal dispersions. This work was financially supported by Solvay. ANR (Agence Nationale de la Recherche) and CGI (Commissariat à l'Investissement d'Avenir) are gratefully acknowledged for their financial support of this work through Labex SEAM (Science and Engineering for Advanced Materials and devices) ANR 11 LABX 086, ANR 11 IDEX 05 02. We acknowledge the ImagoSeine facility (Jacques Monod Institute, Paris, France), and the France BioImaging infrastructure supported by the French National Research Agency (ANR-10-INSB-04, « Investments for the future »). This research was supported in part by the Agence Nationale de la Recherche under the contract ANR-15-CE18-0024-01 (ICONS), ANR-17-CE09-0017 (AlveolusMimics), ANR-20-CE18-0022 (Stric-On) and ANR-21-CE19-0058-1 (MucOnChip).

## Author Contributions

Conceptualization, methodology and investigation, E.K.O.; Formal analysis and writing—original draft preparation, E.K.O. and J.-F.B.; Resources, writing—review and editing and funding acquisition, J.-F.B. All authors have read and agreed to the published version of the manuscript.

## Conflicts of Interest

The authors declare no conflict of interest.

## Graphical abstract

## Fragrance capsules on cellulose

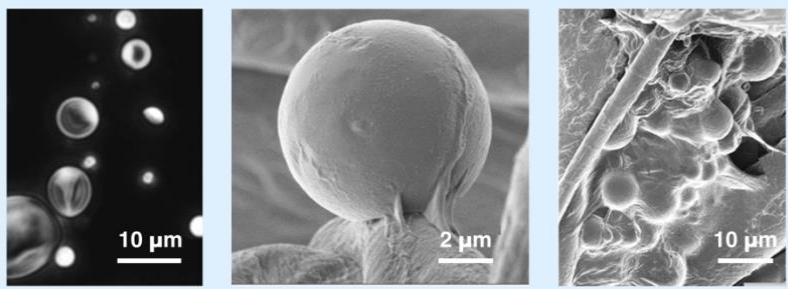